\pdfoutput=1
\documentclass[pra,twocolumn,superscriptaddress,longbibliography]{revtex4-2}
\usepackage{graphicx,empheq,amssymb,scalerel,tikz}
\usepackage[colorlinks=true, citecolor=blue, urlcolor=blue, linkcolor=blue ]{hyperref}
\usetikzlibrary{svg.path}
\definecolor{orcidlogocol}{HTML}{A6CE39}
\tikzset{
	orcidlogo/.pic={
		\fill[orcidlogocol] svg{M256,128c0,70.7-57.3,128-128,128C57.3,256,0,198.7,0,128C0,57.3,57.3,0,128,0C198.7,0,256,57.3,256,128z};
		\fill[white] svg{M86.3,186.2H70.9V79.1h15.4v48.4V186.2z}
		svg{M108.9,79.1h41.6c39.6,0,57,28.3,57,53.6c0,27.5-21.5,53.6-56.8,53.6h-41.8V79.1z M124.3,172.4h24.5c34.9,0,42.9-26.5,42.9-39.7c0-21.5-13.7-39.7-43.7-39.7h-23.7V172.4z}
		svg{M88.7,56.8c0,5.5-4.5,10.1-10.1,10.1c-5.6,0-10.1-4.6-10.1-10.1c0-5.6,4.5-10.1,10.1-10.1C84.2,46.7,88.7,51.3,88.7,56.8z};}}
\newcommand\orcid[1]{\href{https://orcid.org/#1}{\mbox{\scalerel*{\begin{tikzpicture}[yscale=-1,transform shape]\pic{orcidlogo};\end{tikzpicture}}{|}}}}

\begin{document}
\title{Quantum speed limit from a quantum-state-diffusion method}

\author{Wei Wu\orcid{0000-0002-7984-1501}}
\affiliation{Key Laboratory of Quantum Theory and Applications of MoE, Lanzhou University, Lanzhou 730000, China}
\affiliation{Lanzhou Center for Theoretical Physics, Key Laboratory of Theoretical Physics of Gansu Province, Lanzhou University, Lanzhou 730000, China}
\author{Jun-Hong An\orcid{0000-0002-3475-0729}}
\email{anjhong@lzu.edu.cn}
\affiliation{Key Laboratory of Quantum Theory and Applications of MoE, Lanzhou University, Lanzhou 730000, China}
\affiliation{Lanzhou Center for Theoretical Physics, Key Laboratory of Theoretical Physics of Gansu Province, Lanzhou University, Lanzhou 730000, China}

\begin{abstract}
Characterizing the most efficient evolution, the quantum speed limit (QSL) plays a significant role in quantum technology. How to generalize the well-established QSL from closed systems to open systems has attracted much attention. In contrast to the previous schemes to derive the QSL from the reduced dynamics of open system, we propose a QSL bound from the point of view of the total system consisting of the open system and its environment using a quantum-state-diffusion method. The application of our scheme to a two-level system reveals that the system possesses an infinite speedup capacity in the noiseless case, which is destroyed by the environment under the Born-Markovian approximation. It is interesting to find that the capacity in the noiseless case is recovered in the non-Markovian dynamics as long as a bound state is formed in the energy spectrum of the total system. Enriching the characterization schemes of the QSL, our result provides an efficient way to control the QSL of open systems.
\end{abstract}
\maketitle

\section{Introduction}
Quantum mechanics imposes a fundamental limit on the evolution speed of quantum systems, which is called the quantum speed limit (QSL). Mandelstam and Tamm originally found that a lower bound on time taken by a system to evolve to its orthogonal state is determined by the energy variance \cite{JPhys(USSR),PhysRevLett.65.1697,doi:10.1119/1.16940,PhysRevLett.124.110601,PhysRevLett.126.180603,Hornedal_2022} due to the celebrated energy-time uncertainty relation~\cite{PhysRevA.31.2078,Nicholson2020}.  Margolus and Levitin found another bound of the QSL governed by the mean energy relative to the ground state \cite{MARGOLUS1998188}. Sun and Zheng derived a QSL bound relating to the geometric phase via the gauge invariant distance \cite{PhysRevLett.123.180403}. These three bounds are unified into an elegant form for both Hermitian and non-Hermitian systems \cite{PhysRevLett.127.100404}. The QSL is of significance in understanding the performance of various protocols in quantum technology \cite{PhysRevLett.103.160502,PhysRevLett.120.060409,Campbell_2022,Mohan_2022,Aifer_2022}. It determines the computation limits in quantum computing \cite{Lloyd2000,Santos2015,Campbell_2022,Aggarwal_2022}, the ultimate precisions in quantum metrology \cite{Giovannetti2011,PhysRevA.85.052127,Campbell_2018},
and the maximal power in thermodynamic devices, such as quantum engines and batteries \cite{Campo2014,PhysRevLett.118.150601,Campaioli_2022}. It has been experimentally demonstrated in Refs. \cite{Bason2012,Frank2016,doi:10.1126/sciadv.aau5999,doi:10.1126/sciadv.abj9119}.

There is an increasing interest in exploring the QSL in open systems \cite{PhysRevLett.110.050402,PhysRevLett.110.050403,PhysRevLett.111.010402,PhysRevLett.114.233602,PhysRevLett.115.210402,Zhang2014,Sun2015,PhysRevX.6.021031,PhysRevX.12.011038,Deffner_2017,Funo_2019,Teittinen_2019,PhysRevA.103.022210,PhysRevA.103.022221,MONDAL2016689,Nakajima_2022,Lan_2022}. First, people desire to know if the well-established QSL in closed systems is generalizable to open systems. From an application perspective, the QSL of open systems governs how the performance limits of different quantum protocols are impacted by the environment. A Mandelstam-Tamm-type bound on the QSL time for pure initial states was derived by using a completely positive nonunitary map \cite{PhysRevLett.110.050402,PhysRevLett.110.050403}, which was generalized to the non-Markovian dynamics \cite{PhysRevLett.111.010402}. After some debates \cite{Okuyama_2018}, a unified bound on the QSL time including the Mandelstam-Tamm \cite{JPhys(USSR)}, Margolus-Levitin \cite{MARGOLUS1998188}, and Sun-Zheng \cite{PhysRevLett.123.180403} types for time-dependent non-Hermitian systems was established in Ref. \cite{PhysRevLett.127.100404}. A QSL for arbitrary initial states was obtained by introducing the relative purity \cite{Zhang2014} and the Hilbert-Schmidt product of operators \cite{Sun2015} as distance measures. The distinguished roles played by the non-Markovian effect \cite{PhysRevLett.111.010402,PhysRevA.89.012307,PhysRevA.91.032112,PhysRevA.93.020105,Teittinen_2019}, initial entanglement \cite{PhysRevLett.115.210402,PhysRevA.91.022102}, and quantum criticality \cite{Wei2016,PhysRevA.100.022118,PhysRevA.102.053716} in the QSL of open systems have been extensively studied. However, almost all these studies rely on the reduced dynamics of open systems described by master equations. Unfortunately, exact master equations are available only for very few open systems. This dramatically constrains the exploration of the QSL in open systems. The utilization of certain approximations, e.g., the Born-Markovian and secular approximations, or the perturbation methods in the weak-coupling condition, e.g., the Nakajima-Zwanzig method \cite{breuer2002theory}, may be possible solutions, but they might miss important physics. A natural question is if it is possible to derive a QSL bound without \emph{a priori} knowledge of the reduced dynamics of open systems.

We propose a scheme to answer this question by using the quantum-state-diffusion (QSD) method~\cite{PhysRevA.58.1699,PhysRevLett.82.1801,PhysRevA.60.91,PhysRevA.69.052115,PhysRevA.85.032123,PhysRevLett.105.240403,PhysRevLett.119.180401,PhysRevA.98.012110}. Our scheme characterizes the QSL from a global perspective of the total system consisting of the open system and its environment instead of from the reduced density matrix. It is particularly useful for the non-Markovian dynamics, where the system and the environment are highly entangled such that the separation of their degrees of freedom becomes hard. We apply our formulation to a two-level system and reveal its sufficient speedup capacity in the noiseless case. However, such a capacity is destroyed by the Born-Markovian approximate decoherence~\cite{Deffner_2017,PhysRevA.103.022221,PhysRevA.101.042107,PhysRevA.100.052305,Xu_2016}. A mechanism to restore the speedup capacity is found in the non-Markovian dynamics. It is due to the formation of a bound state in the energy spectrum of the total system. This result supplies us an insightful instruction to suppress the destructive effect of the environment on the QSL of the open system.

\section{QSL from a QSD method }
The QSL can be established in a geometric formalism \cite{2017Sebastian,Deffner_2017,PhysRevX.6.021031,PhysRevLett.120.070401,PhysRevX.10.021056,PhysRevResearch.2.023299,PhysRevA.104.052432}. Consider an evolution from $|\psi(0)\rangle$ to $|\psi(\tau)\rangle$, which forms a path in the state space connecting $|\psi(0)\rangle$ and $|\psi(\tau)\rangle$. The length of this path is the line integral $\ell\equiv\int_{0}^{\tau}dt\sqrt{g_{tt}}$, where $g_{tt}$ is a metric determined by the evolution. Thus, the average speed of the evolution is $\bar{v}=\ell/\tau$. However, the path $\ell$ is not the shortest one connecting $|\psi(0)\rangle$ and $|\psi(\tau)\rangle$, which instead is the geodesic with length $\mathcal{L}$. Therefore, $\mathcal{L}$ gives a lower bound for the length of the path between $|\psi(0)\rangle$ and $|\psi(\tau)\rangle$. According to Refs. \cite{PhysRevX.6.021031,Funo_2019,PhysRevLett.123.180403,PhysRevLett.127.100404}, the QSL time is the ratio of the geodesic and the average speed $\bar{v}$, i.e.,
\begin{equation}
\tau_\text{QSL}=\mathcal{L}/\bar{v}=\mathcal{L}\tau/\ell,\label{qsldef}
\end{equation}
which makes a connection between the QSL and the actual evolution path. Physically, $\tau_{\text{QSL}}$ quantifies the smallest evolution time from $|\psi(0)\rangle$ to $|\psi(\tau)\rangle$. By manipulating the actual evolution path as close as possible to the geodesic, we can increase the speedup capacity of the system. When $\tau_{\mathrm{QSL}}=\tau$, the actual evolution path tends to the geodesic and the system generally has no more space for speedup. In contrast, when $\tau_{\text{QSL}}/\tau$ approaches zero, the actual path dramatically deviates from the geodesic and the system has a remarkable speedup capacity. Thus, the smaller $\tau_{\mathrm{QSL}}/\tau$ is, the more speedup capability the system may possess. Such a viewpoint has been widely adopted~\cite{2017Sebastian,PhysRevA.93.020105,Xu2020,PhysRevA.91.022102,PhysRevA.100.052305,PhysRevA.98.022114}, although there still exist some controversies on the physical explanation of $\tau_{\mathrm{QSL}}/\tau$. The geodesic and the metric for a closed system under a unitary evolution are widely chosen as the Bures angle $\mathcal{L}_{\text{B}}=\arccos|\langle\psi(0)|\psi(\tau)\rangle|$ and the Fubini-Study metric $g^\text{FS}_{tt}=\langle\dot{\psi}(t)|\dot{\psi}(t)\rangle-|\langle\dot{\psi}(t)|\psi(t)\rangle|^2$, respectively~\cite{PhysRevLett.65.1697,PhysRevA.36.3479,PhysRevLett.123.180403,PhysRevLett.127.100404}. It is noted that the brachistochrone time of certain system under a time-dependent optimal control may be smaller than the Mandelstam-Tamm-type QSL \cite{PhysRevLett.111.260501,PhysRevLett.114.170501}.

To generalize this characterization from closed systems to open systems, we resort to the QSD method. We study an open system interacting with a bosonic environment. The Hamiltonian reads ($\hbar=1$)\begin{equation}
\hat{H}=\hat{H}_{\text{s}}+\sum_{k}\big[\omega_{k}\hat{b}_{k}^{\dagger}\hat{b}_{k}+\big{(}g_{k}\hat{L}^{\dagger}\hat{b}_{k}+\text{H}.\text{c}.\big{)}\big],\label{hamdt}
\end{equation}
where $\hat{H}_{\text{s}}$ is the system Hamiltonian, $\hat{L}$ is the coupling operator of the system to the environment, $\hat{b}_{k}$ is the annihilation operator of the $k$th environmental mode with frequency $\omega_{k}$, and $g_k$ is the system-environment coupling strength. The coupling is described by the spectral density $J(\omega)\equiv\sum_{k}|g_{k}|^{2}\delta(\omega-\omega_{k})$. We consider an Ohmic spectral density
$J(\omega)=\eta\omega e^{-\omega/\omega_{c}}$, where $\eta$ is a coupling constant and $\omega_{c}$ is a cutoff frequency \cite{RevModPhys.59.1}.

The conventional methods to describe the dynamics of open systems start from the unitary evolution of the total system formed by the system and its environment and trace out the environmental degrees of freedom. It results in a master equation satisfied by the reduced density matrix of the system. Unfortunately, due to the involved infinite degrees of freedom of the environment, this process generally cannot be exactly done and many approximations, e.g., Born-Markovian and secular approximations, are needed. Different from those methods, we use the QSD method. It permits us to follow each quantum trajectory parametrized by the introduced stochastic variable and obtain an exact non-Markovian dynamics without resorting to the reduced density matrix. According to this method, the exact dynamics of the open system is governed by the stochastic Schr\"{o}dinger equation~\cite{PhysRevA.58.1699,PhysRevLett.82.1801,PhysRevA.60.91,PhysRevA.69.052115,PhysRevA.85.032123,PhysRevLett.105.240403,PhysRevLett.119.180401,PhysRevA.98.012110}
\begin{equation}
i|\dot{\psi}_{\bar{\pmb z}}(t)\rangle=\hat{\mathcal{H}}|\psi_{\bar{\pmb z}}(t)\rangle,\label{nhmdyn}
\end{equation}
where $|{\pmb z}\rangle=\bigotimes_ke^{z_k \hat{b}_k^\dag}|0_k\rangle$ is the coherent state of the environment and $|\psi_{\bar{\pmb z}}(t)\rangle=\langle \bar{\pmb z}|\Psi_{\text{T}}(t)\rangle$, with $\bar{\pmb z}$ denoting the complex conjugate of ${\pmb z}$ and $|\Psi_{\text{T}}(t)\rangle$ being the state of the total system. The effective time-dependent non-Hermitian Hamiltonian $\hat{\mathcal{H}}$ is (see Appendix \ref{appmet})
\begin{equation}
\hat{\mathcal{H}}=\hat{H}_{\text{s}}+i\hat{L}\bar{\pmb z}_{t}-i\hat{L}^{\dagger}\mathcal{\hat{O}}(t,\bar{\pmb z}),\label{nhmhmt}
\end{equation}
where $\bar{\pmb z}_{t}=-i\sum_{k}\bar{g}_{k}\bar{z}_{k}e^{i\omega_{k}t}$ can be seen as a stochastic-noise variable and $
\mathcal{\hat{O}}(t,\bar{\pmb z})=\int_{0}^{t}ds\alpha(t-s)\hat{O}(t,s,\bar{\pmb z})$, with $\hat{O}(t,s,\bar{\pmb z})=\frac{\delta}{\delta \bar{\pmb z}_{s}}$ and $\alpha(t-s)=\int_{0}^{\infty}d\omega J(\omega)e^{-i\omega(t-s)}$ being the environmental correlation function. The dynamics in Eq. \eqref{nhmdyn} governed by Eq. \eqref{nhmhmt} defines a single quantum trajectory parametrized by the stochastic variable ${\pmb z}_t$. Defined as $\mathcal{M}\{\mathcal{F}_{\bar{\pmb z},{\pmb z}}\}\equiv\prod_k\int {d^{2}z_k\over \pi}e^{-|z_k|^{2}}\mathcal{F}_{\bar{\pmb z},{\pmb z}}$ for any variable $\mathcal{F}_{\bar{\pmb z},{\pmb z}}$, the ensemble average over all the quantum trajectories recovers the reduced density matrix in the conventional methods. Assuming $|\Psi_{\text{T}}(0)\rangle=|\psi(0)\rangle\bigotimes_{k}|0_{k}\rangle$, one can prove that the variable ${\pmb z}_t$ satisfies $\mathcal{M}\{{\pmb z}_{t}\}=\mathcal{M}\{{\pmb z}_{t}{\pmb z}_{s}\}=0$ and $\mathcal{M}\{{\pmb z}_{t}\bar{\pmb z}_{s}\}=\alpha(t-s)$. The QSD method has been widely used in studying quantum dissipative dynamics~\cite{PhysRevLett.105.240403,PhysRevLett.119.180401,PhysRevA.98.012110,PhysRevE.84.051112,PhysRevLett.120.150402,PhysRevA.97.042126,PhysRevA.98.032116}. $\mathcal{\hat{O}}(t,\bar{\pmb{z}})$ is analytically solvable for a multilevel system~\cite{PhysRevLett.105.240403} and the quantum Brownian-motion model~\cite{PhysRevA.69.052115}. In general situations, $\mathcal{\hat{O}}(t,\bar{\pmb{z}})$ is perturbatively~\cite{PhysRevA.60.91,Xu_2014} or numerically~\cite{PhysRevA.97.042126,PhysRevA.98.032116,PhysRevLett.113.150403} computable. As long as $\mathcal{\hat{O}}(t,\bar{\pmb{z}})$ is known, the dynamics of the open system is obtained. Note that the QSD method is an exact description of the dynamics of the open system with the quantum-jump terms self-consistently included.

Regarding the open system and the environment as a total system, we can safely apply $\mathcal{L}_{\text{B}}$ and $g_{tt}^{\text{FS}}$ in the pure state $|\Psi_\text{T}(\tau)\rangle$ as $\mathcal{L}_{\text{B}}=\arccos|\langle\Psi_{\text{T}}(0)|\Psi_{\text{T}}(\tau)\rangle|$ and $g_{tt}^{\text{FS}}=\langle\dot{\Psi}_{\text{T}}(t)|\dot{\Psi}_{\text{T}}(t)\rangle-|\langle\dot{\Psi}_{\text{T}}(t)|\Psi_{\text{T}}(t)\rangle|^2$. Then, by inserting $\frac{1}{\pi}\int d^{2}\pmb{z} e^{-|\pmb{z}|^{2}}|\pmb{z}\rangle\langle\pmb{z}|=\pmb{1}$, we obtain
\begin{eqnarray}
\mathcal{L}_{\text{B}}&=&\arccos|\mathcal{M}\{\langle\psi_{\pmb{z}}(0)|\psi_{\bar{\pmb{z}}}(\tau)\rangle\}|,\label{eq:eqlb}\\
g^\text{FS}_{tt} &=&\mathcal{M}\{\langle \dot{\psi}_{{\pmb z}}(t)|\dot{\psi}_{\bar{\pmb z}}(t)\rangle\}-|\mathcal{M}\{\langle \dot{\psi}_{{\pmb z}}(t)|\psi_{\bar{\pmb z}}(t)\rangle\}|^2.\label{eq:eqgtt}
\end{eqnarray}
The environmental effects have been incorporated into the nonunitary evolution and the ensemble average to all the quantum trajectories. Equation \eqref{eq:eqgtt} is recast into $g^\text{FS}_{tt} =\mathcal{M}\{\langle\psi_{{\pmb z}}(t)|\hat{\mathcal H}^{\dagger}\hat{\mathcal H}|\psi_{\bar{\pmb z}}(t)\rangle\}-|\mathcal{M}\{\langle\psi_{{\pmb z}}(t)|\hat{\mathcal H}|\psi_{\bar{\pmb z}}(t)\rangle\}|^2$ by using Eq. \eqref{nhmdyn}, which returns to the energy variance revealed by the Mandelstam-Tamm bound in the noiseless case \cite{JPhys(USSR),PhysRevLett.65.1697,doi:10.1119/1.16940,PhysRevLett.124.110601,PhysRevLett.126.180603}. Thus, we have succeeded in converting the characterization of the QSL of an open system into the well-established one of a closed system. Giving a global picture of the QSL of an open system from the total system, our scheme depicts a different facet of the QSL of an open system from the conventional methods. The similar idea of characterizing the physical properties of an open system from the joint system-environment state has been widely employed in the studies of geometric phase~\cite{PhysRevA.71.044101,PhysRevA.89.062118}, non-Markovianity~\cite{Pernice_2012,PhysRevA.98.012142}, and QSL~\cite{PhysRevLett.110.050402}.

\section{Exemplification}
\begin{figure}
\centering
\includegraphics[width=\columnwidth]{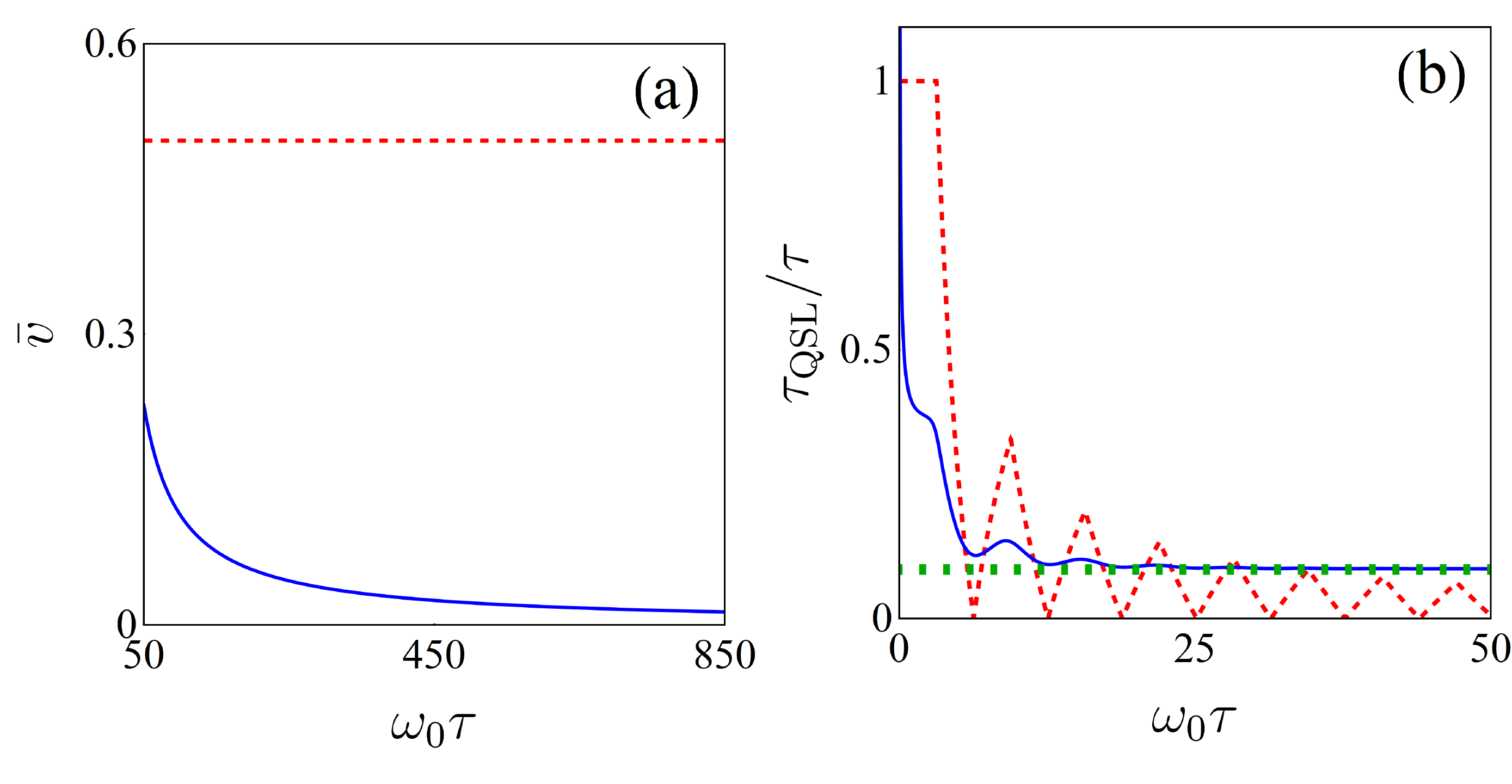}
\caption{Average speed $\bar{v}$ (a) and QSL time $\tau_{\text{QSL}}/\tau$ (b) as a function of $\omega_{0}\tau$. The red dashed and blue solid lines are the results in the ideal case and the Born-Markovian approximate case, respectively. The green dotted line is the analytical long-time result $\lim_{\tau\rightarrow\infty}\tau_\text{QSL}/\tau=\pi\kappa/(3A)$. We use $\eta=0.05$ and $\omega_c=10\omega_0$. }\label{fig:fig1}
\end{figure}

To verify the feasibility of our scheme, we consider a two-level system with $\hat{H}_{\text{s}}=\omega_{0}\hat{\sigma}_{+}\hat{\sigma}_{-}$, where $\hat{\sigma}_\pm$ are the transition operators between the ground state $|g\rangle$ and excited state $|e\rangle$. In the noiseless case, one calculates $|\psi(\tau)\rangle =e^{-i\hat{H}_{\text{s}}\tau}|\psi(0)\rangle$ from a general initial state $|\psi(0)\rangle=\sin(\theta/ 2)|e\rangle+\cos(\theta/ 2)e^{i\phi}|g\rangle$. It leads to $g_{tt}^\text{FS}=\omega_0^2\sin^2\theta/4$ and $\mathcal{L}_{\text{B}}=\arccos[\sqrt{3+\cos(2\theta)+[1-\cos(2\theta)]\cos(\omega_0\tau)}/2]$. The maximal average speed reads $\bar{v}=\omega_0/2$, as confirmed by Fig. \ref{fig:fig1}(a). The maximal QSL time $\tau^\text{ide}_{\text{QSL}}=2\omega_{0}^{-1}\arccos|\cos(\omega_{0}\tau/2)|$ is achieved according to Eq. \eqref{qsldef} when $\theta=\pi/2$ \cite{PhysRevLett.111.260501}. We find that, after keeping in one within a short time duration, $\tau^\text{ide}_{\text{QSL}}/\tau$ tends to zero with increasing $\tau$, see Fig. \ref{fig:fig1}(b). This implies that the system has a sufficient speedup capacity in the noiseless case.

In the presence of an environment, we choose $\hat{L}=\hat{\sigma}_{-}$, which is widely used to verify the performance of different QSL bounds \cite{PhysRevLett.110.050402,PhysRevLett.111.010402,PhysRevA.93.020105,PhysRevX.6.021031}. We can derive $\mathcal{\hat{O}}(t)=\hat{\sigma}_{-}\int_{0}^{t}ds\alpha(t-s)\frac{u(s)}{u(t)}$ and $u(t)$ satisfying (see Appendix \ref{appmet})
\begin{equation}
\dot{u}(t)+i\omega_{0} u(t)+\int_{0}^{t}ds\alpha(t-s)u(s)=0,\label{EOM}
\end{equation}under $u(0)=1$. Then the Bures angle and the Fubini-Study metric are calculated as (see Appendix \ref{appqal})
\begin{eqnarray}
\mathcal{L}_{\text{B}}&=&\arccos[{|1+u(\tau)| }/{2}],\label{exmll}\\
g_{tt}^{\text{FS}}&=&[|\dot{u}(t)|^2+\alpha(0)|u(t)|^2]/2\nonumber\\
&&-\big|i[\bar{u}(t)\dot{u}(t)-u(t)\dot{\bar u}(t)]-\omega_0|u(t)|^2\big|^2/4.\label{exmgg}
\end{eqnarray}
We consider the QSL of the system relaxing to its steady state by choosing a sufficiently large $\tau$, which reflects the equilibration efficiency of the open system. The QSL derived in such a condition is an important factor in characterizing the decoherence.

In the special case when the system-environment coupling is weak and the typical time scale of the environment is much smaller than the one of the system, we can apply the Born-Markovian approximation to Eq. \eqref{EOM}. Its approximate solution reads $u(t)=e^{-[\kappa+i(\omega_{0}+\Delta_{\omega_{0}})]t}$~\cite{PhysRevE.90.022122}, where $\kappa=\pi J(\omega_{0})$ is the decay rate and $\Delta_{\omega_{0}}=\mathcal{P}\int _0^\infty d\omega{J(\omega)\over \omega_{0}-\omega}$ is a frequency shift. Substituting it into Eqs. \eqref{exmll} and \eqref{exmgg}, we have $\mathcal{L}_{\text{B,BMA}}=\arccos[|1+e^{-2\kappa \tau}+2e^{-\kappa\tau}\cos(\omega_0\tau+\Delta_{\omega_{0}}\tau)|/2]$ and $g_{tt,\text{BMA}}^{\text{FS}}\simeq A^2e^{-2\kappa t}$, with $A^2=[\alpha(0)+\kappa^2+(\omega_0+\Delta_{\omega_{0}})^2]/2$. Then we obtain $\lim_{\tau\rightarrow \infty}\bar{v}=0$ and $\lim_{\tau\rightarrow\infty}\tau_\text{QSL}/\tau=\pi\kappa/(3A)$ (see Appendix \ref{appqal}). Thus, the speedup capacity is destroyed by the environment under this approximation, which is in agreement with previous results \cite{PhysRevA.103.022221,PhysRevA.101.042107,PhysRevA.100.052305,Xu_2016,Deffner_2017}. The Born-Markovian approximate $\bar{v}$ and $\tau_\text{QSL}/\tau$ in Fig. \ref{fig:fig1} verify our analytical conclusion that $\bar{v}$ tends to zero and $\tau_\text{QS}/\tau$ tends to $\pi\kappa/(3A)$ asymptotically.

In the general non-Markovian dynamics, the expressions of $\bar{v}$ and $\tau_{\text{QSL}}/\tau$ are complicated. We leave them to numerical calculations. However, via analyzing the long-time behavior of $u(t)$, we can obtain their asymptotic forms, which are helpful to us in building up a physical picture of the QSL of the open system. We apply a Laplace transform to Eq. \eqref{EOM} and find
$\tilde{u}(z)\equiv\int_{0}^{\infty}dt u(t)e^{-zt}=[z+i\omega_{0}+\int_0^\infty d\omega{J(\omega)\over z+i\omega}]^{-1}$. The solution of $u(\tau)$ is obtained by applying an inverse Laplace
transform to $\tilde{u}(z)$, which can be done by finding the poles from
\begin{equation}~\label{eq:eq13}
y(\varpi)\equiv\omega_{0}-\int_0^\infty d\omega{J(\omega)\over\omega-\varpi} =\varpi,~(\varpi=iz).
\end{equation}
It is remarkable to find that the root of Eq. \eqref{eq:eq13} is the eigenenergy of $\hat{H}$ in the single-excitation subspace. To be more specific, we express the eigenstate as $|\Phi\rangle=(x\hat{\sigma}_{+}+\sum_{k}y_{k}\hat{b}_{k}^{\dagger})|0,\{0_k\}\rangle$. Substituting it into $\hat{H}|\Phi\rangle=E|\Phi\rangle$, one finds the eigenenergy equation as $E-\omega_{0}-\sum_{k}g_{k}^{2}/(\omega_{k}-E)=0$, which matches Eq.~(\ref{eq:eq13}) in the continuous limit of the environmental frequency after replacing $E$ by $\varpi$. This implies that the dynamics of the open system is essentially determined by the energy-spectrum features of the total system. $y(\varpi)$ is a monotonically decreasing function when $\varpi<0$. Thus, Eq.~(\ref{eq:eq13}) has one isolated root $E_b$ in this regime provided $y(0)<0$. Because $y(\varpi)$ is not well defined in the regime $\varpi>0$ due to the poles in its integrand, Eq.~(\ref{eq:eq13}) has an infinite number of roots in this regime, which form a continuous energy band. We call the eigenstate corresponding to the isolated eigenenergy $E_b$ the bound state~\cite{PhysRevA.81.052330,PhysRevA.87.052139,PhysRevA.103.L010601}. Using the residue theorem, we have
\begin{equation}
u(t)=Ze^{-iE_b t}+\int_{0}^{\infty}\frac{J(\omega)e^{-i\omega t}d\omega}{(\omega-\omega_{0}-\Delta_\omega)^{2}+[2\pi J(\omega)]^2},\nonumber
\end{equation}
where the first term with $Z\equiv[1+\int_0^\infty{J(\omega)d\omega\over(E_b-\omega)^2}]^{-1}$ is contributed by the bound state. The second term is from the energy-band states and approaches zero in the long-time regime due to out-of-phase interference. Thus, if the bound state is absent, we have $u(\infty)=0$ implying a complete decoherence, while if the bound state is formed, we have $u(\infty)=Ze^{-iE_bt}$ implying a decoherence suppression. It is determined from $y(0)< 0$ that the bound state is formed when $\omega_0<\eta\omega_c $ for the Ohmic spectral density.

\begin{figure}
\centering
\includegraphics[width=\columnwidth]{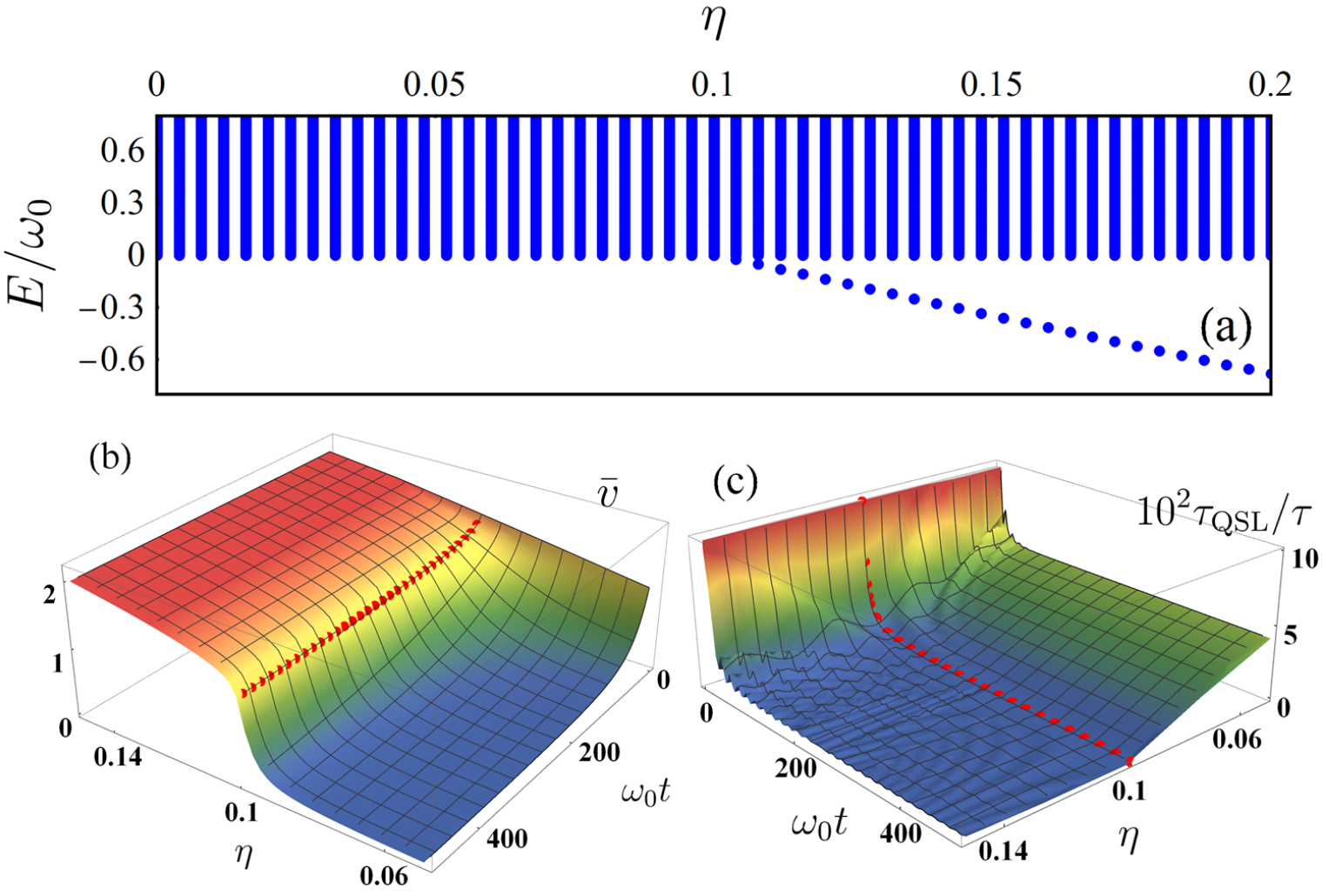}
\caption{(a) Energy spectrum of the total system. Exact non-Markovian result of the average speed (b) and the QSL time (c) in different $\eta$ and $\omega_0\tau$. The dashed lines highlight the behaviors at the critical point $\eta=0.1$ for forming the bound state. Parameter values are the same as Fig. \ref{fig:fig1}. }\label{fig:fig2}
\end{figure}

In the absence of the bound state, it is natural to expect that the QSL has a consistent behavior with the one under the Born-Markovian approximation because $u(t)$ approaches zero eventually. We focus on the case in the presence of the bound state. Using the form $u(\infty)=Ze^{-iE_bt}$, we obtain $\lim_{\tau\rightarrow\infty}\bar{v}\simeq C$ and
\begin{equation}
\lim_{\tau\rightarrow\infty}{\tau_\text{QSL}\over\tau}={\arccos[\sqrt{1+Z^2+2Z\cos(E_b\tau)}/2]\over C\tau},\label{limcod}
\end{equation}
where $C^2=Z^2[\alpha(0)+E_b^2]/2-Z^4(\omega_0/2-E_b)^2$~(see Appendix \ref{appqal}). It recovers the ideal result $\tau^\text{ide}_{\text{QSL}}/\tau$ in the limit of $g_k$ tending to zero, which reduces $E_b=\omega_0$ and $Z=1$. Equation \eqref{limcod} tends to zero in the long-time limit, which is consistent with the result in the ideal case. This interesting result reveals that, quite different from the Born-Markovian approximate result in Fig. \ref{fig:fig1}, the formation of the bound state recovers the speedup capacity of the open system.

\begin{figure}
\centering
\includegraphics[width=\columnwidth]{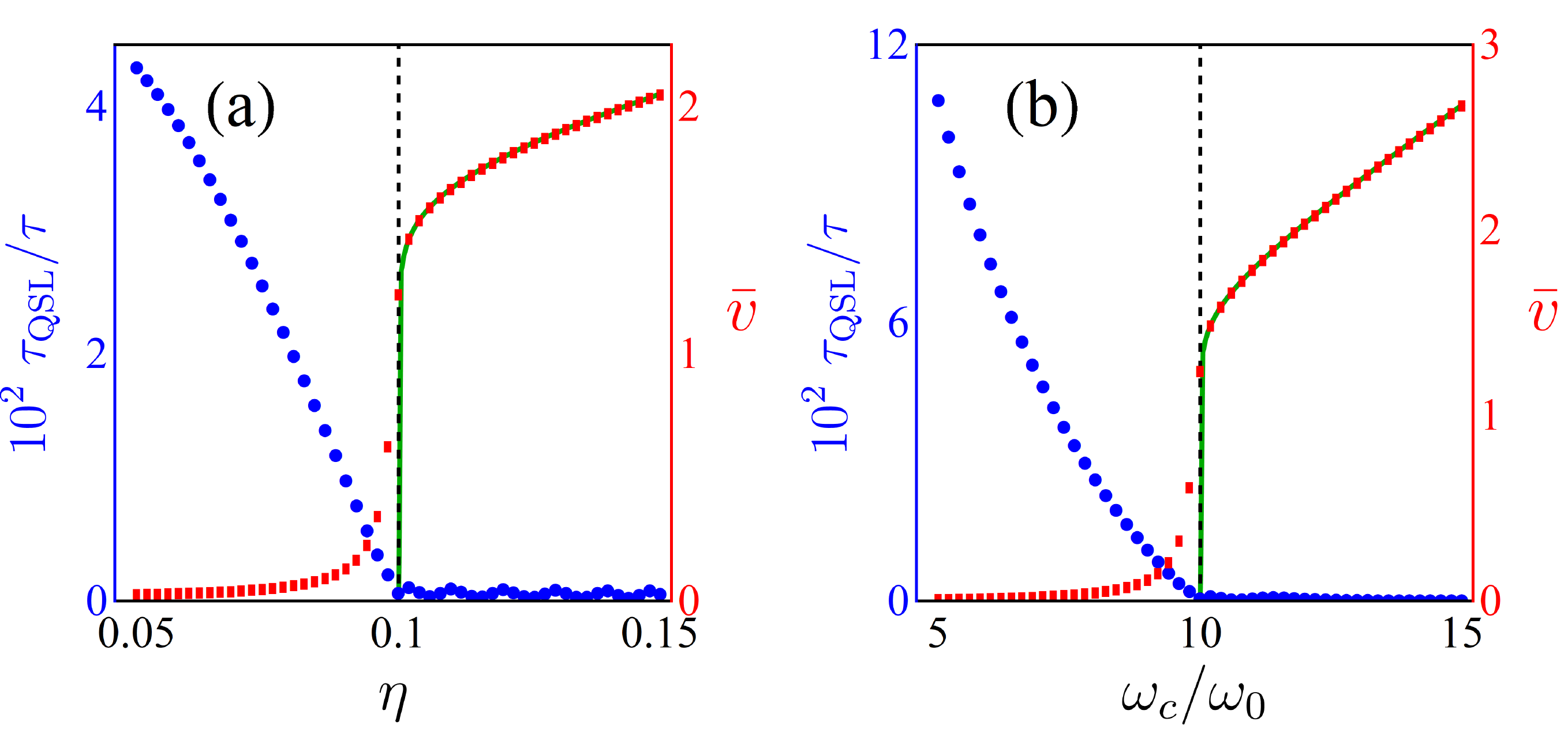}
\caption{(a) Steady-state average speed (red rectangles) and QSL time (blue circles) as a function of $\eta$ (a) and $\omega_c/\omega_0$ (b) when $\omega_0\tau=800$. The green solid lines are from the analytical result $
\lim_{\tau\rightarrow\infty}\bar{v}=C$. Parameter values are the same as Fig. \ref{fig:fig1}. }\label{fig:fig3}
\end{figure}

We plot in Fig. \ref{fig:fig2}(a) the energy spectrum of the total system consisting of the two-level system and its environment. It is found that the energy branch of the bound state splits the energy spectrum into two regions, without the bound state when $\eta<0.1$ and with the bound state when $\eta>0.1$. An obvious threshold at the critical point $\eta=0.1$ for forming the bound state occurs both for the average speed $\bar{v}$ and the QSL time $\tau_\text{QSL}/\tau$, see Figs. \ref{fig:fig2}(b) and \ref{fig:fig2}(c). When $\eta<0.1$, $\bar{v}$ tends to zero and $\tau_\text{QSL}/\tau$ tends to a finite value. Thus the speedup capacity of the system is completely destroyed by the environment. This result is qualitatively similar to the one under the Born-Markovian approximation in Fig. \ref{fig:fig1}. In contrast, when $\eta>0.1$, $\bar{v}$ tends to a finite constant and $\tau_\text{QSL}/\tau$ tends to zero in the long-time limit. It indicates the ideal speedup capacity of the system is recovered due to the formation of the bound state. Figure \ref{fig:fig3} shows the long-time behaviors of $\bar{v}$ and $\tau_\text{QSL}/\tau$ in different $\eta$ and $\omega_c$. It confirms that as long as the bound state of the total system is formed, the speedup capacity of the open system is restored.

Our result is generalizable to the spin-boson model after relaxing the rotating-wave approximation. It is interesting to find that the distinguished role played by the bound state in restoring the speedup capacity still works (see Appendix \ref{appsbms}). Further, although only the two-level system is studied, our scheme can be readily applied in the continuous-variable systems, where a similar dominated role played the bound state in determining the decoherence dynamics exists \cite{PhysRevA.103.L010601,PhysRevApplied.17.034073}. These indicate the generality of the insight of the QSL of open systems gained by our scheme.

\section{Discussions and conclusion}
Our result implies that we can control the QSL and manipulate the speedup capacity of the open system via engineering the formation of the bound state, which can be realized by the quantum reservoir engineering technique \cite{ER1,Kienzler53,PhysRevA.78.010101}. It is noted that, although only the Ohmic spectral density is considered, our bound-state mechanism is applicable to other forms, where the explicit condition for forming the bound state may be different, but the bound-state mechanism does not change. The bound state and its dynamical effect have been experimentally observed in circuit QED~\cite{Liu2016} and ultracold atom~\cite{Kri2018,Kwon2022} systems. These experimental progresses provide strong support to test our characterization scheme of the QSL and to verify our mechanism of recovering the speedup capacity of open systems.

As a global-picture characterization of the QSL of the open system from the total system formed by the system and its environment, our proposed QSL bound reflects different facets of the QSL of open system from the one derived from the reduced density matrix. In Appendix \ref{appdrelt}, we discuss the relation between our proposed QSL bound and the previous bound based on the reduced density matrix. On the other hand, we only use the Fubini-Study metric as an example to illustrate our method of deriving the QSL bound. The validity of our method and the obtained conclusion does not depend on the specific form of the used metric.

In summary, we have proposed a scheme to characterize the QSL of open systems using the QSD method. Without resorting to the reduced density matrix, it is a global reflection of the QSL of an open system from the total system consisting of the open system and its environment. The application in a two-level system reveals that the system has a remarkable speedup capacity in the ideal case, which is fully destroyed by the Born-Markovian approximate decoherence. A mechanism to retrieve the capacity by engineering the formation of a bound state in the energy spectrum of the total system has been found in the non-Markovian dynamics. Our result enriches the characterization scheme of the QSL and supplies an insightful guideline to control the QSL of open systems. It might attract experimental interests in controlling decoherence and evaluating the ultimate performance of various quantum protocols.

\section{Acknowledgments}
This work is supported by the National Natural Science Foundation of China (Grants No. 12275109, No. 11834005, and No. 12247101).

\appendix

\section{QSD method}\label{appmet}

We consider an open system interacting with an environment governed by Eq. \eqref{hamdt}. The Schr\"{o}dinger equation in the rotating frame of $\hat{H}_\text{e}$ is
\begin{equation}
i|\dot{\Psi}_\text{T}(t)\rangle=\big[\hat{H}_\text{s}+\sum_k\big(g_k\hat{L}\hat{b}^\dag_ke^{i\omega_k t}+\text{H.c.}\big)\big]|\Psi_\text{T}\rangle.\label{smseq}
\end{equation}
In the representation formed by the coherent state $|{\pmb z}\rangle=\bigotimes_k |z_k\rangle$ with $|z_k\rangle\equiv e^{z_k\hat{b}_k^\dag}|0_k\rangle$, Eq. \eqref{smseq} reads
\begin{equation}
i|\dot{\psi}_{\bar{\pmb z}}(t)\rangle=\big[\hat{H}_\text{s}+i\hat{L}{\bar{\pmb z}}_t+\hat{L}^\dag\sum_kg_k^*e^{-i\omega_kt}\partial_{\bar{z}_k}\big]|\psi_{\bar{\pmb z}}(t)\rangle,\label{smcsb}
\end{equation}
where $|\psi_{\bar{\pmb z}}(t)\rangle=\langle {\bar{\pmb z}}|\Psi_\text{T}(t)\rangle $, ${\bar{\pmb z}}_t=-i\sum_k g_k\bar{z}_k e^{i\omega_k t}$, $\langle \bar{z}_k|\hat{b}_k^\dag=\bar{z}_k\langle\bar{z}_k|$, and $\langle \bar{z}_k|\hat{b}_k=\partial_{\bar{z}_k}\langle\bar{z}_k|$ have been used  \cite{PhysRevA.58.1699,PhysRevLett.82.1801}. The chain rule $\partial_{\bar{z}_k}=\int_0^tds{\partial\bar{\pmb z}_s\over\partial \bar{z}_k}{\delta\over \delta \bar{\pmb z}_s}=-i g_k \int_0^tds e^{i\omega_k s}{\delta\over \delta \bar{\pmb z}_s}$ converts Eq. \eqref{smcsb} into
\begin{eqnarray}
&&i|\dot{\psi}_{\bar{\pmb z}}(t)\rangle=\hat{\mathcal H}|\psi_{\bar{\pmb z}}(t)\rangle,\label{smcsb2}\\
&&\hat{\mathcal H}=\hat{H}_\text{s}+i\hat{L}{\bar{\pmb z}}_t-i\hat{L}^\dag\mathcal{\hat{O}}(t,\bar{\pmb z}),\label{smnhh}
\end{eqnarray}
where $\alpha(t-s)=\sum_k|g_k|^2 e^{-i\omega_k(t-s)}$ is the environmental correlation function and $\mathcal{\hat{O}}(t,\bar{\pmb z})\equiv\int_{0}^{t}ds\alpha(t-s)\hat{O}(t,s,\bar{\pmb z})$, with $\hat{O}(t,s,\bar{\pmb z})=\frac{\delta}{\delta \bar{\pmb z}_{s}}$. Equation \eqref{smcsb2} can be seen as a stochastic Schr\"{o}dinger equation governed by a non-Hermitian Hamiltonian \eqref{smnhh}. As long as $\mathcal{\hat{O}}(t,\bar{\pmb z})$ is determined, the dynamics of the open system is obtained.

The form of $\hat{O}(t,s,\bar{\pmb z})$ for a two-level system with $\hat{H}_\text{s}=\omega_0\hat{\sigma}_+\hat{\sigma}_-$ and $\hat{L}=\hat{\sigma}_-$ must be ~\cite{PhysRevA.69.052115,PhysRevLett.119.180401}
\begin{equation}
\hat{O}(t,s,\bar{\pmb z})|\psi_{\bar{\pmb z}}(t)\rangle= f(t,s)\hat{\sigma}_-|\psi_{\bar{\pmb z}}(t)\rangle,\label{smddd}
\end{equation}where $f(t,s)$ is under determined, because the term with $\hat{\sigma}_+$ multiplying $\hat{L}^\dag=\hat{\sigma}_+$ equals to zero. Differentiating Eq. \eqref{smddd} with respect to $t$, we obtain
\begin{equation}
\dot{f}(t,s)\hat{\sigma}_-|\psi_{\bar{\pmb z}}(t)\rangle=\big{[}\frac{\delta}{\delta \bar{\pmb z}_{s}}-f(t,s)\hat{\sigma}_-\big{]}|\dot{\psi}_{\bar{\pmb z}}(t)\rangle.\label{smsf}
\end{equation}
The substitution of Eqs. \eqref{smcsb2} and \eqref{smddd} into Eq. \eqref{smsf} results in
\begin{eqnarray}
\dot{f}(t,s)&=&f(t,s)[i\omega_0+\int_0^tds'\alpha(t-s')f(t,s')].\label{smfeq}
\end{eqnarray}
Setting $f(t,s)=u(s)/u(t)$, we rewrite Eq. \eqref{smfeq} as
\begin{equation}
\dot{u}(t)+i\omega_0u(t)+\int_0^t\alpha(t-s)u(s)ds=0.\label{smuu}
\end{equation}
Therefore, we find
\begin{eqnarray}
\mathcal{\hat{O}}(t,\bar{\pmb z})=\int_{0}^{t}ds\alpha(t-s){u(s)\over u(t)}\hat{\sigma}_-\equiv F(t)\hat{\sigma}_-,\label{smo}
\end{eqnarray}with $F(t)=\int_0^tds \alpha(t-s)u(s)/u(t)$.

Next, we prove that Eq.~(\ref{smcsb2}) recovers the exact master equation after making the average over $\pmb{\bar{z}}_{t}$.
Substituting Eqs. \eqref{smo} and \eqref{smnhh} into Eq. \eqref{smcsb2}, we obtain
\begin{equation}
|\dot{\psi}_{\bar{\pmb z}}(t)\rangle=[-i\omega_{0}\hat{\sigma}_{+}\hat{\sigma}_{-}+\bar{\pmb z}_{t}\hat{\sigma}_{-}-F(t)\hat{\sigma}_{+}\hat{\sigma}_{-}]|\psi_{\bar{\pmb z}}(t)\rangle.
\end{equation}
The equation of motion of $\rho(t)=\mathcal{M}\{|\psi_{\pmb{\bar{z}}}(t)\rangle\langle\psi_{\pmb{z}}(t)|\}$ is
\begin{eqnarray}
\dot{\rho}(t)&=&-i[\hat{H}_{\text{s}},\rho(t)]-F(t)\hat{\sigma}_{+}\hat{\sigma}_{-}\rho(t)-F^{*}(t)\rho(t)\hat{\sigma}_{+}\hat{\sigma}_{-}\nonumber\\
&&+[\hat{\sigma}_{-}\mathcal{M}\{\pmb{\bar{z}}_{t}|\psi_{\pmb{\bar{z}}}(t)\rangle\langle\psi_{\pmb{z}}(t)|\}+\text{H.c.}].~~~~~~~
\end{eqnarray}
According to the Furutsu-Novikov theorem
$\mathcal{M}\{\pmb{\bar{z}}_{t}\mathcal{P}_{t}\}=\int_{0}^{t}ds \mathcal{M}\{\pmb{\bar{z}}_{t}\pmb{z}_{s}\}\mathcal{M}\big{\{}\frac{\delta\mathcal{P}_{t}}{\delta \pmb{z}_{s}}\big{\}}$ ~\cite{PhysRevA.58.1699,PhysRevLett.82.1801,PhysRevA.98.012110},
with $\mathcal{P}_{t}=|\psi_{\pmb{\bar{z}}}(t)\rangle\langle\psi_{\pmb{z}}(t)|$ being the projection operator, we have
\begin{eqnarray}
\mathcal{M}\{\pmb{\bar{z}}_{t}|\psi_{\pmb{\bar{z}}}(t)\rangle\langle\psi_{\pmb{z}}(t)|\}&=&F^{*}(t)\rho(t)\hat{\sigma}_{+}.
\end{eqnarray}
Then we obtain
\begin{equation}
\begin{split}
\dot{\rho}(t)=&i\text{Im}\bigg{[}\frac{\dot{u}(t)}{u(t)}\bigg{]}[\hat{\sigma}_{+}\hat{\sigma}_{-},\rho(t)]-\text{Re}\bigg{[}\frac{\dot{u}(t)}{u(t)}\bigg{]}\\
&\times[2\hat{\sigma}_{-}\rho(t)\hat{\sigma}_{+}-\hat{\sigma}_{+}\hat{\sigma}_{-}\rho(t)-\rho(t)\hat{\sigma}_{+}\hat{\sigma}_{-}],
\end{split}
\end{equation}
where $F(t)=-i\omega_{0}-\dot{u}(t)/u(t)$ from Eq. \eqref{smuu} has been used. It is just the non-Markovian master equation \cite{Breuer}.

\section{QSL for a dissipative two-level system}\label{appqal}

\begin{figure}[tbp]
\centering
\includegraphics[angle=0,width=8.8cm]{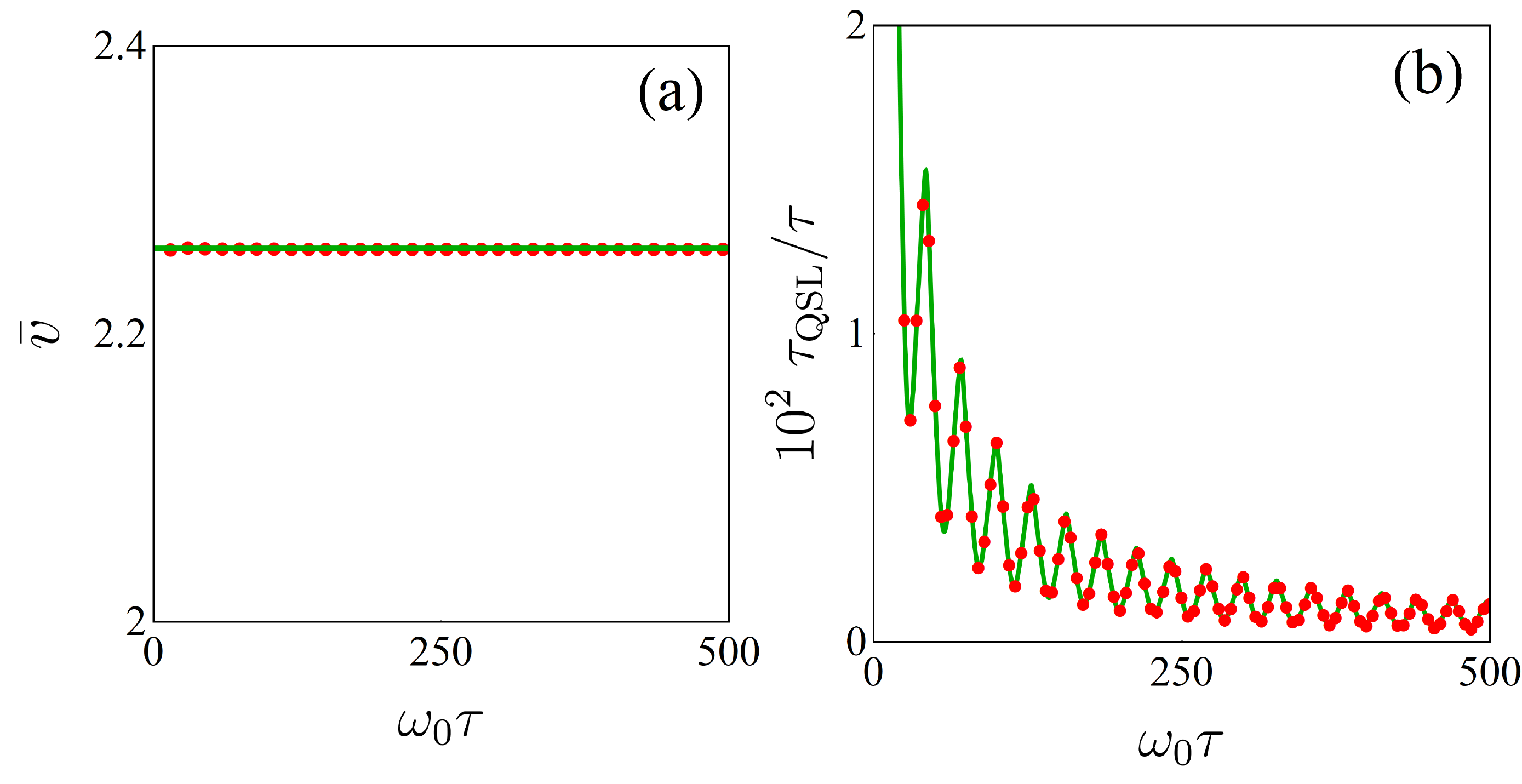}
\caption{Numerical results (green solid lines) of $\bar{v}$ (a) and $\tau_\text{QSL}/\tau$ (b) as a  function of $\omega_0\tau$ when $\omega_c=13\omega_0$. The red dots are the results evaluated by Eqs. \eqref{smvnm} and \eqref{smtanm}, respectively. We use $\eta=0.1$.  }\label{fig:figsm}
\end{figure}
We derive the QSL time for a dissipative two-level system. Equation \eqref{smnhh} reads $\hat{\mathcal H}=i\dot{u}(t)/u(t)\hat{\sigma}_+\hat{\sigma}_-+i\bar{\pmb z}_{t}\hat{\sigma}_{-}$.
Consider that the initial state is $|\psi_{\bar{\pmb z}}(0)\rangle=(|g\rangle+|e\rangle)/\sqrt{2}$. Its evolution is expanded as $|\psi_{\bar{\pmb z}}(t)\rangle=c_g(t)|g\rangle+c_e(t)|e\rangle$. The substitution of this state into Eq. \eqref{smcsb2} results in
\begin{eqnarray}
c_e(t)=\frac{u(t)}{\sqrt{2}},~~c_g(t)=\frac{1}{\sqrt{2}}+\frac{1}{\sqrt{2}}\int_{0}^{t}ds\bar{\pmb z}_{s}u(s).
\end{eqnarray}
Then, the geodesic length is given by
\begin{eqnarray}
\mathcal{L}_{\text{B}}&=&\arccos\frac{\big{|\mathcal{M}\{c_{g}(\tau)+c_{e}(\tau)\}\big{|}}}{\sqrt{2}}=\arccos\frac{|1+u(\tau)| }{2},~~~~~\label{smlll}
\end{eqnarray}
where $\mathcal{M}\{\bar{\pmb{z}}_{s}\}=0$ has been used. In the limit of $g_k$ approaching zero, we have $u^{\text{ideal}}(\tau)= e^{-i\omega_{0}\tau}$ and thus $\mathcal{L}_{\text{B}}^{\text{ideal}}=\arccos|\cos(\frac{1}{2}\omega_{0}\tau)|$. Similarly, one obtains
\begin{eqnarray}
g_{tt}^{\text{FS}}&=&[|\dot{u}(t)|^2+\alpha(0)|u(t)|^2]/2\nonumber\\
&&-\big|i[\bar{u}(t)\dot{u}(t)-u(t)\dot{\bar u}(t)]-\omega_0|u(t)|^2\big|^2/4,\label{smgtt}
\end{eqnarray}
with reduces to $\frac{1}{4}\omega_{0}^{2}$ in the ideal limit.

In the Born-Markovian approximation, the solution of $u(t)$ reads $u_{\text{BMA}}(t)\simeq e^{-\{\kappa+i[\omega_{0}+\Delta(\omega_{0})]\}t}$. Substituting $u_\text{BMA}(t)$ into Eqs. \eqref{smlll} and \eqref{smgtt}, we obtain
\begin{eqnarray}
\mathcal{L}_{\text{B,BMA}}&=&\arccos[|1+e^{-2\kappa \tau}+2e^{-\kappa\tau}\nonumber\\
&&\times\cos(\omega_0\tau+\Delta_{\omega_{0}}\tau)|/2],\label{smlb}\\
g_{tt,\text{BMA}}^{\text{FS}}&=&{ e^{-2\kappa t}\over 2}[\alpha(0)+\kappa^2+(\omega_0+\Delta_{\omega_0})^2]\nonumber\\
&&-{e^{-4\kappa t}\over 4}(\omega_0+2\Delta_{\omega_0})^2\simeq A^2e^{-2\kappa t},
\end{eqnarray}
where $A^2=[\alpha(0)+\kappa^2+(\omega_0+\Delta_{\omega_0})^2]/2$.
Thus, the average speed $\bar{v}$ is calculated as
\begin{equation}
\bar{v}_{\text{BMA}}=\frac{1}{\tau}\int_{0}^{\tau}dt\sqrt{g_{tt}^{\text{FS}}}=\frac{A(1-e^{-\kappa\tau})}{\kappa\tau},\label{smbv}
\end{equation}
which vanishes in the long-time limit. The QSL time is
\begin{equation}
\lim_{\tau\rightarrow\infty}\frac{\tau_{\text{QSL,BMA}}}{\tau}=\frac{\pi\kappa}{3A},
\end{equation}
which is a constant.

In the non-Markovian dynamics, we focus on the case in the presence of the bound state. The substitution of $u(\infty)=Ze^{-iE_b\tau}$ into Eqs. \eqref{smlll} and \eqref{smgtt} results in
\begin{eqnarray}
\lim_{\tau\rightarrow\infty}\mathcal{L}_\text{B}&=&\arccos{|1+Ze^{-iE_b\tau}|\over 2},\\
\lim_{t\rightarrow\infty}g_{tt}^{\text{FS}}&=&C^2
\end{eqnarray}with $C^2=Z^2[\alpha(0)+E_b^2]/2-Z^4(\omega_0/2-E_b)^2$. We obtain
\begin{eqnarray}
\lim_{\tau\rightarrow\infty}\bar{v}&=&C,\label{smvnm}\\
\lim_{\tau\rightarrow\infty}\frac{\tau_{\text{QSL}}}{\tau}&=&{\arccos[\sqrt{1+Z^2+2Z\cos(E_b\tau)}/2]\over C\tau}.~~~~~\label{smtanm}
\end{eqnarray}
We plot in Fig.~\ref{fig:figsm} the exact results by numerically solving Eq.~\eqref{smuu} and the analytical results in Eqs.~\eqref{smvnm} and \eqref{smtanm}. We see that the analytical results are in qualitative agreement with those of the numerical results.

\section{QSL for the spin-boson model}\label{appsbms}

\begin{figure}[tbp]
\centering
\includegraphics[angle=0,width=8.0cm]{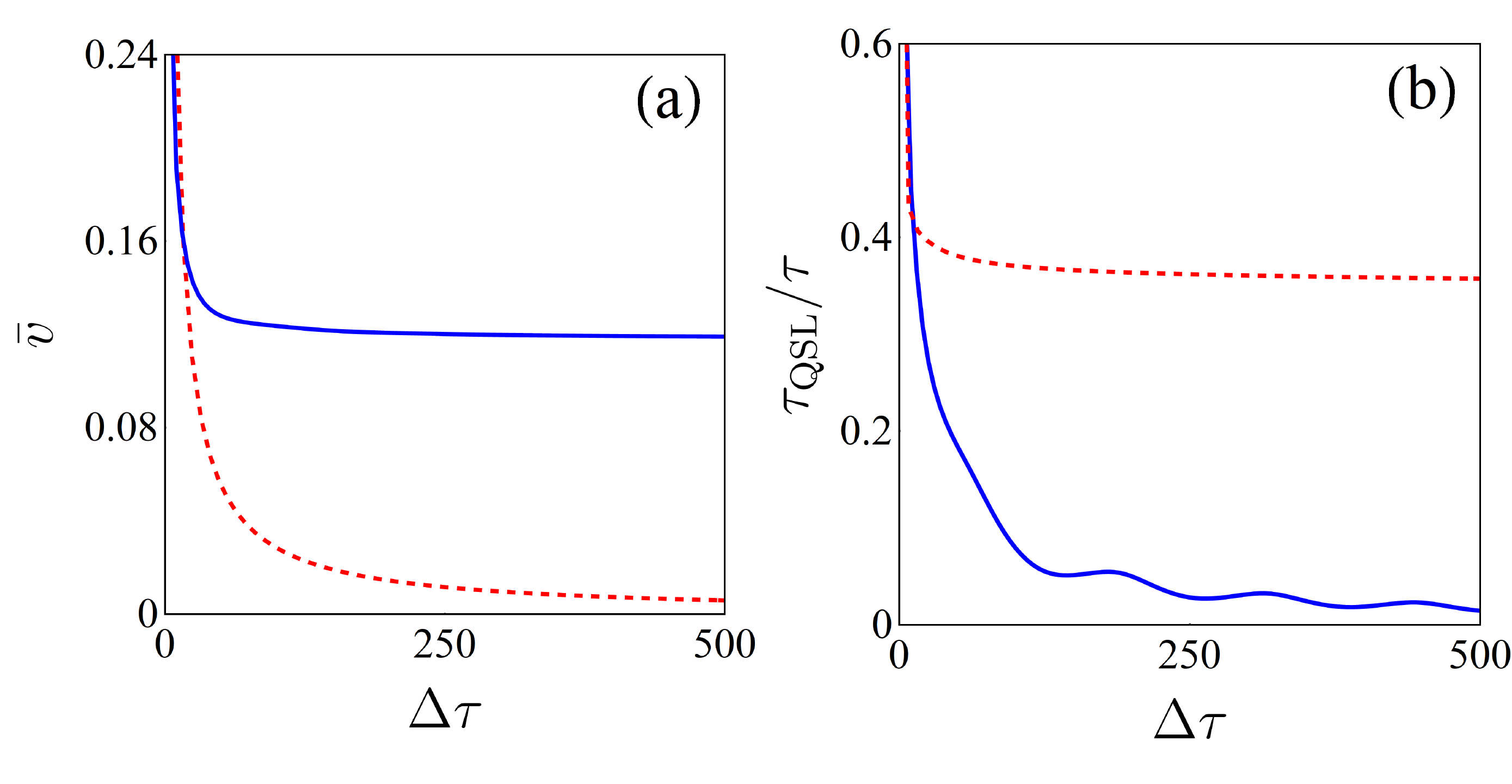}
\caption{Numerical results of $\bar{v}$ (a) and $\tau_\text{QSL}/\tau$ (b) as a function of $\omega_0\tau$ for the spin-boson model. The blue solid lines are results in the presence of the bound state when $\eta=0.25$. The red dashed lines are the results in the absence of the bound state when $\eta=0.1$. The spectral density is chosen as $J(\omega)=\eta\omega^{s}\omega_{c}^{1-s}e^{-\omega/\omega_{c}}$ with $s=0.6$ and $\omega_{c}=30\Delta$.}\label{fig:figsm2}
\end{figure}

\begin{figure}[tbp]
\centering
\includegraphics[angle=0,width=8.8cm]{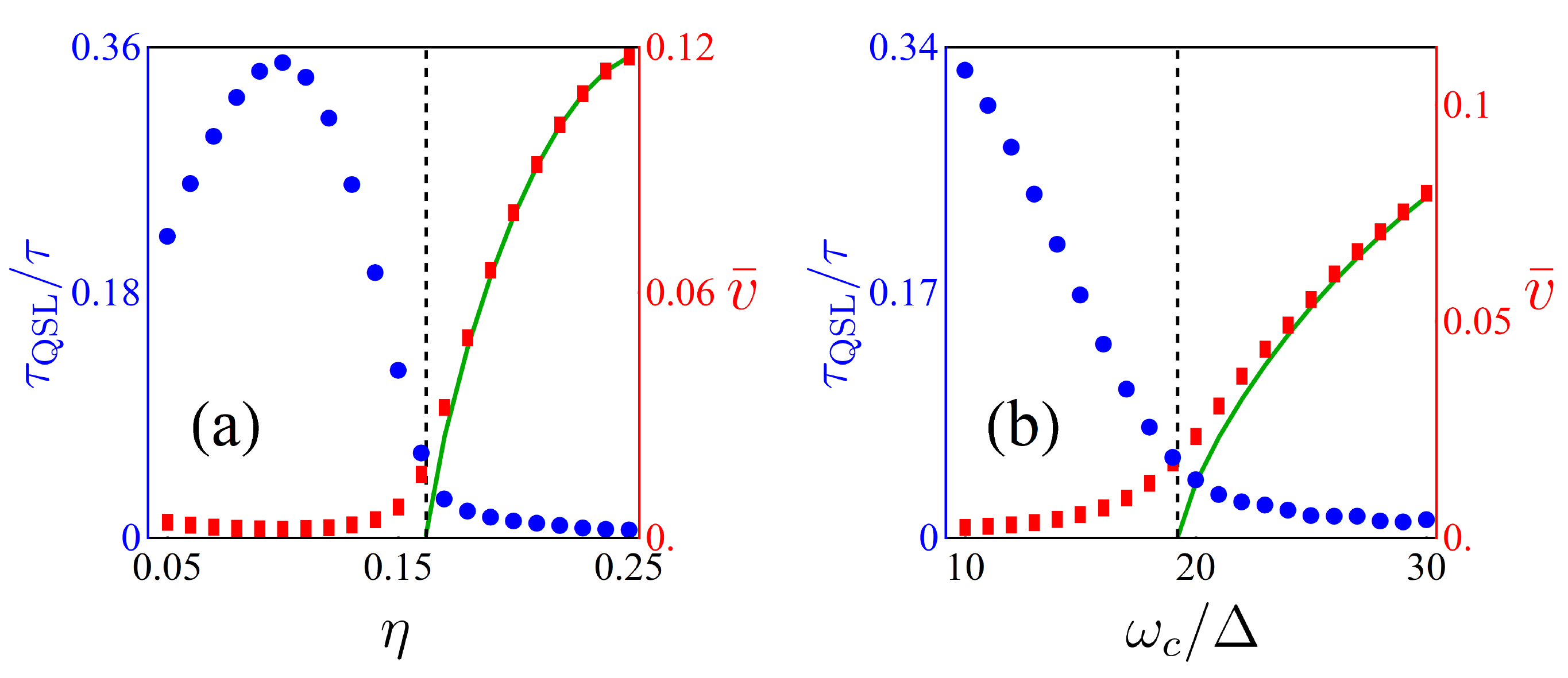}
\caption{Steady-state $\bar{v}$ (red rectangles) and $\tau_\text{QSL}/\tau$ (blue circles) as a function of $\eta$ with $\omega_{c}=30\Delta$ (a) and $\omega_{c}/\Delta$ with $\eta=0.2$ (b) when $\Delta\tau=10^3$. The green solid lines are obtained from the analytical results by using $\tilde{u}(\infty)\simeq \tilde{Z}e^{-i\tilde{E}_{b}t}$. The black dashed lines mark the critical points of forming the bound state. Other parameters are the same as Fig.~\ref{fig:figsm2}.}\label{fig:figsm3}
\end{figure}

For the spin-boson model, i.e., $\hat{H}_{\text{s}}=\frac{1}{2}\Delta\hat{\sigma}_{z}$ and $\hat{L}=\frac{1}{2}\hat{\sigma}_{x}$, an exact QSL is not obtainable due to the complexity of the model itself. An approximate form can be derived in the weak-coupling condition. We apply a polaron transformation $e^{\hat{S}}$, with $\hat{S}=\exp[\sum_{k}\frac{g_{k}\xi_{k}}{2\omega_{k}}(\hat{b}_{k}^{\dagger}-\hat{b}_{k})\hat{\sigma}_{x}]$, and obtain $\hat{H}'=\hat{H}_{0}'+\hat{V}'$ ~\cite{PhysRevB.75.054302,PhysRevB.84.174301,doi:10.1063/1.4722336}, where
\begin{eqnarray}
\hat{H}_{0}'&=&\frac{1}{2}\Delta\Theta\hat{\sigma}_{z}+\sum_{k}\omega_{k}\hat{b}_{k}^{\dagger}\hat{b}_{k}+\sum_{k}\frac{g_{k}^{2}}{4\omega_{k}}\xi_{k}(\xi_{k}-2),~~~\\
\hat{V}'&=&\frac{1}{2}\hat{\sigma}_{x}\sum_{k}g_{k}(1-\xi_{k})(\hat{b}_{k}^{\dagger}+\hat{b}_{k})-\frac{i}{2}\Delta\hat{\sigma}_{y}\sinh\hat{\chi}\nonumber \\
&&+\frac{1}{2}\Delta\hat{\sigma}_{z}(\cosh\hat{\chi}-\Theta),
\end{eqnarray}
with $\hat{\chi}=\sum_{k}\frac{g_{k}\xi_{k}}{\omega_{k}}(\hat{b}_{k}^{\dagger}-\hat{b}_{k})$ and
$\Theta=\langle\{0_{k}\}|\cosh\hat{\chi}|\{0_{k}\}\rangle=\exp\big{(}-\sum_{k}\frac{g_{k}^{2}\xi_{k}^{2}}{2\omega_{k}^{2}}\big{)}$.
The value of $\xi_{k}$ is determined by minimizing the Bogoliubov-Feynman upper bound of the free energy $
F_{\text{B}}=-\frac{1}{\beta}\ln\big{(}\text{Tr}e^{-\beta\hat{H}_{0}'}\big{)}+\langle \hat{V}'\rangle_{\hat{H}_{0}'}$,
where $\langle \hat{V}'\rangle_{\hat{H}_{0}'}$ is the thermal expectation value with respect to the Gibbs state of $\hat{H}_{0}'$. Thus, one obtains $\xi_{k}=\frac{\omega_{k}}{\omega_{k}+\Theta\Delta}$.
Then, $\hat{V}'$ is rewritten as $\hat{V}'=\hat{V}'_{1}+\hat{V}'_{2}$ with
\begin{eqnarray}
\hat{V}'_{1}&=&\sum_{k}\tilde{g}_{k}(\hat{b}_{k}\hat{\sigma}_{+}+\hat{b}_{k}^{\dagger}\hat{\sigma}_{-}),\\
\hat{V}'_{2}&=&\frac{1}{2}\Delta\hat{\sigma}_{z}(\cosh\hat{\chi}-\Theta)-\frac{i}{2}\Delta\hat{\sigma}_{y}(\sinh\hat{\chi}-\Theta\hat{\chi}),~~~
\end{eqnarray}
where $\hat{\sigma}_{\pm}=(\hat{\sigma}_{x}\pm i\hat{\sigma}_{y})/2$ and $\tilde{g}_{k}=g_{k}\Theta\Delta/(\omega_{k}+\Theta\Delta)$ is the renormalized coupling strength. It is observed that $\hat{V}'_{2}$ is the multiboson transition of the order $O(\tilde{g}_{k}^{4})$. Such multiboson processes can be neglected if the system-environment coupling is weak. Thus, we finally have
\begin{eqnarray}
\hat{H}'_{\text{eff}}=\Theta\Delta\hat{\sigma}_{+}\hat{\sigma}_{-}+\sum_{k}[\omega_{k}\hat{b}_{k}^{\dagger}\hat{b}_{k}+\tilde{g}_{k}(\hat{b}_{k}\hat{\sigma}_{+}+\text{H.c.})],~~\label{smsmbh}
\end{eqnarray}
where the constant term has been dropped. We see that Eq. \eqref{smsmbh} has the same structure as Eq. \eqref{hamdt}, with the system frequency and the coupling strength renormalized by the polaron transformation.

Substituting Eq. \eqref{smsmbh} into \eqref{smnhh}, we obtain
\begin{eqnarray}
\mathcal{\hat{H}}_{\text{SBM}}&=&\Theta\Delta\hat{\sigma}_{+}\hat{\sigma}_{-}+i\bar{\pmb{z}}_{t}\hat{\sigma}_{-}\nonumber \\
&&-i\int_{0}^{t}ds\tilde{\alpha}(t-s)\frac{\tilde{u}(s)}{\tilde{u}(t)}\hat{\sigma}_{+}\hat{\sigma}_{-},
\end{eqnarray}
where $\tilde{\alpha}(t)=\sum_{k}|\tilde{g}_{k}|^{2}e^{-i\omega_{k}t}$ and $\tilde{u}(t)$ is determined by
\begin{equation}
\dot{\tilde{u}}(t)+i\Theta\Delta \tilde{u}(t)+\int_{0}^{t}ds\tilde{\alpha}(t-s)\tilde{u}(s)=0.
\end{equation}
A similar analysis obtains that a bound state with eigenenergy $\tilde{E}_{b}$ is formed provided $\tilde{y}(0)<0$, where
\begin{equation}
\tilde{y}(\varpi)=\Theta\Delta-\int_{0}^{\infty}d\omega\frac{J(\omega)}{\omega-\varpi}\bigg{(}\frac{\Theta\Delta}{\Theta\Delta+\omega}\bigg{)}^{2}.
\end{equation}
Then, $\tilde{u}(\infty)\simeq \tilde{Z}e^{-i\tilde{E}_{b}t}$ with $\tilde{Z}=[1+\int_{0}^{\infty}d\omega\frac{J(\omega)}{(\tilde{E}_{b}-\omega)^{2}}(\frac{\Theta\Delta}{\Theta\Delta+\omega})^{2}]^{-1}$. In contrast, if $\tilde{y}(0)\geq0$, no bound state is formed and $\tilde{u}(\infty)=0$. With $\hat{\mathcal{H}}_{\text{SBM}}$ at hand, the QSL for the spin-boson model is readily obtained by using the same method as the main text. As plotted in Figs.~\ref{fig:figsm2} and \ref{fig:figsm3}, one can conclude that the bound-state mechanism to retrieve the ideal speedup capacity still works for the spin-boson model.

\section{Relation with previous QSL bounds}\label{appdrelt}

We discuss the relation between our proposed QSL bound and the previous bound obtained from the reduced density matrix of the open system. The Fubini-Study metric is closely related to the quantum Fisher information as
\begin{equation}
\begin{split}
g_{tt}^{\text{FS}}=\frac{1}{4}\mathcal{F}(t).
\end{split}
\end{equation}
Here, $\mathcal{F}(t)=4[\langle\dot{\Psi}_{\text{T}}(t)|\dot{\Psi}_{\text{T}}(t)\rangle-|\langle\dot{\Psi}_{\text{T}}(t)|\Psi_{\text{T}}(t)\rangle|^{2}]$ is nothing else but the quantum Fisher information with respect to the state of the total system. Then, based on the fact that the quantum Fisher information of the total system is an upper bound to that of the subsystem~\cite{PhysRevLett.110.050402}, we have
\begin{equation}
\mathcal{F}(t)\geq\mathcal{F}_{\text{red}}(t),
\end{equation}
where $\mathcal{F}_{\text{red}}(t)$ is the quantum Fisher information corresponding to the reduced density matrix $\rho(t)$. Using these results, one finds 
\begin{eqnarray}
\ell&=&\int_{0}^{\tau}dt\sqrt{\frac{1}{4}\mathcal{F}(t)}\geq\int_{0}^{\tau}dt\sqrt{\frac{1}{4}\mathcal{F}_{\text{red}}(t)}=\ell_{\text{red}}.
\end{eqnarray} 
This inequality means that our proposed average speed $\bar{v}=\ell/\tau$ is faster than the one derived from the reduced density matrix $\bar{v}_{\text{red}}=\ell_{\text{red}}/\tau$.

On the other hand, the Bures angle connecting the initial state $\rho(0)=|\psi(0)\rangle\langle\psi(0)|$ and the evolved state $\rho(\tau)=\text{Tr}_{\text{E}}[|\Psi_{\text{T}}(\tau)\rangle\langle\Psi_{\text{T}}(\tau)|]=\mathcal{M}\{|\psi_{\pmb{\bar{z}}}(\tau)\rangle\langle\psi_{\pmb{z}}(\tau)|\}$ of the subsystem reads $\mathcal{L}_{\text{B}}^{\text{red}}=\arccos\sqrt{F[\rho(0),\rho(\tau)]}$, where $F[\rho(0),\rho(\tau)]=\langle\psi(0)|\rho(\tau)|\psi(0)\rangle$ is the so-called quantum fidelity. Due to the facts that $|\psi_{\pmb{\bar{z}}}(0)\rangle=\langle \pmb{\bar{z}}|\Psi_{\text{tot}}(0)\rangle=\otimes_{k}\langle0_{k}|e^{\bar{z}_{k}\hat{b}_{k}}|0_{k}\rangle|\psi(0)\rangle=|\psi(0)\rangle$ and $0\leq e^{-|\pmb{z}|^{2}}/\pi\leq 1$, we find
\begin{equation}
\begin{split}
F[\rho(0),\rho(\tau)]=&\langle\psi(0)|\mathcal{M}\{|\psi_{\pmb{\bar{z}}}(\tau)\rangle\langle\psi_{\pmb{z}}(\tau)|\}|\psi(0)\rangle\\
=&\frac{1}{\pi}\int d^{2}\pmb{z}e^{-|\pmb{z}|^{2}}|\langle\psi(0)|\psi_{\pmb{\bar{z}}}(\tau)\rangle|^{2}\\
=&\frac{1}{\pi}\int d^{2}\pmb{z}e^{-|\pmb{z}|^{2}}|\langle\psi_{\pmb{z}}(0)|\psi_{\pmb{\bar{z}}}(\tau)\rangle|^{2}\\
\geq&\frac{1}{\pi^{2}}\int d^{2}\pmb{z}e^{-2|\pmb{z}|^{2}}|\langle\psi_{\pmb{z}}(0)|\psi_{\pmb{\bar{z}}}(\tau)\rangle|^{2}.
\end{split}
\end{equation}
Using the Cauchy-Schwarz inequality
\begin{equation}
\bigg{|}\int dx A(x)\bigg{|}^{2}\leq\int dx |A(x)|^{2},
\end{equation}
we have
\begin{equation}
\begin{split}
F[\rho(0),\rho(\tau)]\geq&\bigg{|}\int {d^{2}\pmb{z}\over \pi}e^{-|\pmb{z}|^{2}}\langle\psi_{\pmb{z}}(0)|\psi_{\pmb{\bar{z}}}(\tau)\rangle\bigg{|}^{2}\\
=&|\mathcal{M}\{\langle\psi_{\pmb{z}}(0)|\psi_{\pmb{\bar{z}}}(\tau)\}|^{2}.
\end{split}
\end{equation}
Considering the fact that $\arccos\sqrt x$ is a decreasing function in the region $x\in[0,1]$, the relation between $\mathcal{L}_{\text{B}}^{\text{red}}$ and $\mathcal{L}_{\text{B}}$ can be built as
\begin{eqnarray}
\mathcal{L}_{\text{B}}^{\text{red}}&=&\arccos\sqrt{F[\rho(0),\rho(\tau)]}\nonumber\\
&\leq&\arccos\sqrt{|\mathcal{M}\{\langle\psi_{\pmb{z}}(0)|\psi_{\pmb{z}}(\tau)\}|^{2}}\nonumber\\
&=&\arccos|\mathcal{M}\{\langle\psi_{\pmb{z}}(0)|\psi_{\pmb{\bar{z}}}(\tau)\rangle\}|=\mathcal{L}_{\text{B}}.\label{smdsd2}
\end{eqnarray}
For the two-level system example considered in the main text, the validity of Eq.~(\ref{smdsd2}) is quite obvious 
\begin{equation}
\begin{split}
\mathcal{L}_{\text{B}}^{\text{red}}=&\arccos\sqrt{F[\rho(0),\rho(\tau)]}\\
=&\arccos\sqrt{\frac{1}{4}[1+2\text{Re}u(\tau)+1]}\\
\leq&\arccos\sqrt{\frac{1}{4}[1+2\text{Re}u(\tau)+|u(\tau)|^{2}]}\\
=&\arccos\frac{|1+u(\tau)|}{2}=\mathcal{L}_{\text{B}}.
\end{split}
\end{equation}

We finally prove that $\bar{v}\geq\bar{v}_{\text{red}}$ and $\mathcal{L}_{\text{B}}\geq\mathcal{L}_{\text{B}}^{\text{red}}$. Thus, beyond specific models, it is still difficult to compare the tightness between our proposed bound $\tau_{\text{QSL}}=\mathcal{L}_{\text{B}}/\bar{v}$ and the previous bound from the reduced density matrix $\tau_{\text{QSL}}^{\text{red}}=\mathcal{L}_{\text{B}}^{\text{red}}/\bar{v}_{\text{red}}$ in a universal way. However, motivated by Ref.~\cite{PhysRevLett.110.050402}, these two QSL bounds can be further tightened by introducing a hybrid bound as
\begin{equation}
\tau_{\text{QSL}}^{\text{hybrid}}=\frac{\mathcal{L}_{\text{B}}}{\bar{v}_{\text{red}}}.
\end{equation}

\bibliography{reference.bib}

\end{document}